\documentclass[sigconf, nonacm]{acmart}

\newcommand\vldbdoi{XX.XX/XXX.XX}

\newcommand\vldbavailabilityurl{URL_TO_YOUR_ARTIFACTS}

\usepackage{xcolor}
\usepackage{subcaption}
\usepackage{multirow}
\usepackage{listings}
\usepackage{algpseudocode}
\usepackage{enumitem}

\usepackage[linesnumbered,ruled,vlined]{algorithm2e} 

\begin{document}
\title{Generating Query Context for Relational Databases}

\author{Alekh Jindal,
Jyoti Pandey,
Christina Pavlopoulou,
Ronith PR,
Sharath Prakash,
Shi Qiao,
Shivani Tripathi,
Wangda Zhang}
\email{research@tursio.ai}
\affiliation{%
  \vspace{0.4cm}
  \institution{Tursio}
  \city{Bellevue}
  \country{USA}
  \vspace{0.4cm}
}

\begin{abstract}
Relational databases are the systems of record for business applications, and there is growing demand to query them through natural language interfaces. A key challenge is that AI models need appropriate \emph{query context}, i.e., sample questions paired with their corresponding data-model fragments, to generate accurate SQL. Today, creating this context is a manual, time-consuming process that requires expertise in both SQL and the database schema, leading to a cold start problem for new databases and an ongoing maintenance burden as schemas and query patterns evolve.

We present an automated approach for generating query context for relational databases. Our method defines \emph{query flows} that capture common patterns of data retrieval and analysis questions, systematically generates \emph{data models} by traversing these flows, and instantiates them with specific values sampled from the database using weighted strategies that maximize diversity and coverage. The resulting question--data-model pairs can be used to guide natural language interfaces in accurately querying relational databases. We report on our experience deploying this approach across more than 50 real-world databases connected to Tursio.
\end{abstract}

\maketitle



\section{Introduction}

Relational databases serve as the systems of record for business applications, capturing operational workflows and accumulating massive amounts of structured data over time. To make this data accessible beyond SQL-proficient users, there is growing interest in natural language interfaces powered by generative AI~\cite{snowflake-cortex-analyst, databricks-ai-bi-genie, microsoft-fabric-copilot, thoughtspot-spotter, vanna-ai-training}. These interfaces translate natural language questions into SQL queries, democratizing data-driven decision-making and unlocking insights previously inaccessible to non-technical users. This need is further amplified by the rise of AI-powered ``vibe coding'' tools---such as Cursor~\cite{cursor-ai}, Lovable~\cite{lovable-ai}, Bolt~\cite{bolt-ai}, and Replit~\cite{replit-ai}---that enable users to build full-stack applications through natural language prompts. As these tools generate database-backed applications, they too must query the underlying relational data.

However, a key challenge with these interfaces is that AI models need appropriate {\it query context} to generate accurate and reliable SQL. While schema and metadata information provide a starting point, they are often insufficient on their own~\cite{bird_bench, beaver_benchmark}. To bridge this gap, modern tools rely on sample questions paired with their corresponding SQL queries to teach the model how to query a specific database.
For instance, Databricks Genie~\cite{databricks-ai-bi-genie} accepts sample SQL queries and functions that demonstrate how to use the available data for common questions. Likewise, Snowflake Cortex~\cite{snowflake-cortex-analyst} provides a Verified Query Repository (VQR)---a curated collection of question--SQL pairs that the system references when answering similar questions. Other tools offer analogous mechanisms~\cite{thoughtspot-spotter, vanna-ai-training}.

Creating these sample question--SQL pairs, however, is a manual and time-consuming process that requires expertise in both SQL and the underlying database schema. This creates a {\it cold start problem}, especially for large databases with complex schemas. The challenge extends beyond initial setup: users must also maintain these samples over time as the database and query patterns evolve. Adding or removing samples can shift their semantic ranking and inadvertently affect the accuracy of {\it all} questions. Deciding on the right number of samples is itself difficult---too few may leave query patterns uncovered, while too many can reduce generalization. Some systems further compound this by restricting the number of samples (e.g., Databricks limits Genie spaces to 100 sample questions), making it critical to choose samples that maximize coverage within tight constraints.

\begin{figure}[!t]
  \includegraphics[width=0.475\textwidth]{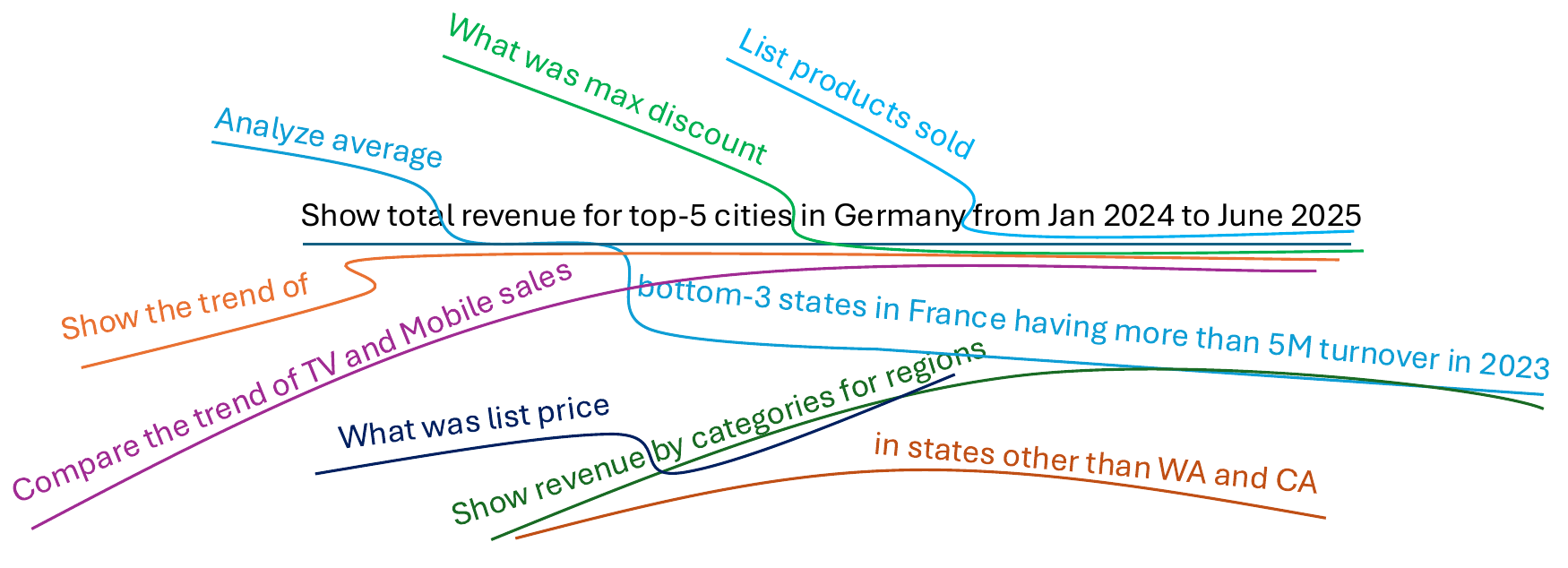}
  \caption{Sample query context to cover different schema elements and data points of a relational database.}
  \Description{Covering data and schema elements}
  \label{fig:question-variations}
  \vspace{-0.2cm}
\end{figure}

In this paper, we present an automated approach for generating query context for relational databases. The goal is to produce a set of question--SQL pairs that systematically cover both the schema elements and the data points of a database. Figure~\ref{fig:question-variations} illustrates this with different variations of the question {\it ``Show total revenue for top-5 cities in Germany from Jan 2024 to June 2025''}, each targeting distinct schema elements and data values. Our key idea is to generate a diverse and representative set of {\it data models}, each being a canonical query tree built bottom-up using relational operators (project, select, join, aggregate, order, limit, etc.). We leverage a combination of query-pattern templates, database schema analysis, and data-sampling techniques to construct these trees and translate them into equivalent natural language questions. The resulting question--data-model pairs can then serve as the query context needed for natural language interfaces to accurately query relational databases.

We make the following contributions:
\begin{enumerate}
    \item A formalization of {\it query context}---a question paired with the data-model fragments needed to answer it---as the mechanism for teaching LLMs how to accurately query relational databases, grounded in decades of query-by-example research (\S\ref{sec:background}).
    \item Algorithms for automatically generating query context by constructing canonical query trees using a combination of query-pattern templates, schema analysis, and data-sampling techniques (\S\ref{sec:approach}).
    \item Deployment experience from over 50 real-world databases, characterizing the scale, runtime, and practical lessons of generating query context automatically (\S\ref{sec:deployment}).
\end{enumerate}

\section{Background and Overview}
\label{sec:background}

We begin by reviewing how examples have long been used to help non-programmers query relational databases, then describe our earlier work on building a {\it large data model} for a database, and finally motivate the need for a high-quality {\it query context} that goes beyond data models.

\subsection{Query-by-Example}

Helping non-programmers query relational databases has been an area of interest for over five decades. The earliest line of work is {\it query-by-example} (QBE), introduced by Zloof~\cite{DBLP:conf/vldb/Zloof75,5388055}, where users fill in table skeletons with example values and the system infers the intended query. This idea has been revisited and extended over the years: Shen et al.~\cite{10.1145/2588555.2593664} presented methods to infer minimal project-join SQL queries that include given example output tuples, while Li et al.~\cite{10.14778/2831360.2831369} proposed an iterative feedback loop with example input/output pairs to progressively converge on the user's intended SQL query. Lissandrini et al.~\cite{exampleBasedMethods} provide a comprehensive survey of example-based data exploration methods.

The common thread across this body of work is that {\it examples} are a powerful mechanism for expressing queries without writing SQL. With LLMs now increasingly responsible for generating SQL in modern applications, a natural question arises: how can we extend the idea of query-by-example to provide the right examples to an LLM, so that it can accurately generate SQL for a given database?

\subsection{Large Data Model}

Raw relational data rarely answers a user's question directly---it typically needs to be transformed through joins, aggregations, filters, and projections into a more useful {\it data model}. The space of such transformations for a given database can be vast, covering different combinations of tables, columns, and aggregation functions. In prior work~\cite{ai-machine}, we addressed this by building a {\it large data model}: a system that enumerates a wide variety of logical data models for a given relational database, each covering a different combination of schema elements and data points. These data models are indexed using vector embeddings and made searchable via natural language queries, allowing users to find the most relevant data model for their question. The matched data model can then be refined into an accurate SQL query.

While the large data model provides broad coverage of data transformations, it does not directly teach LLMs {\it how} to map natural language query patterns to those transformations. This motivates the need for a complementary mechanism---query context---that we describe next.

\subsection{Query Context}

Query context pairs a natural language question with the data-model fragments needed to answer it. By providing a diverse set of such pairs, we can teach LLMs the mapping between how users phrase questions and how the database should be queried. To make this concrete, consider the following natural language question on a healthcare database:

\begin{quotation}
{\it Show lung cancer patients with tumor size greater than 4 who are on lisinopril.}
\end{quotation}

\noindent The relevant query context for this question would include the following data-model fragments, each identifying a specific filter condition:

\begin{itemize}
  \item DIAGNOSIS\_DESCRIPTION = `lung cancer'
  \item TEST = `Tumor size'
  \item TEST\_RESULT $> 4$
  \item DRUG\_NAME = `lisinopril'
\end{itemize}

\noindent With this context, an LLM can understand that ``lung cancer patients'' maps to a filter on the \texttt{DIAGNOSIS\_DESCRIPTION} column, ``tumor size greater than 4'' maps to a comparison on \texttt{TEST\_RESULT}, and so on. Without such context, the LLM would have to guess which columns and values correspond to each part of the question---a common source of errors in text-to-SQL systems~\cite{bird_bench}.

Query context is not limited to filter conditions. Consider a different question on a financial database:

\begin{quotation}
{\it What are the most common cities among the borrowers in the portfolio?}
\end{quotation}

\noindent Here, the query context captures a richer set of relational operations:

\begin{itemize}
  \item Table: PORTFOLIO
  \item Projection: CITY
  \item Aggregation: SUM(number\_of\_records)
  \item Sorting: DESC on total number\_of\_records
\end{itemize}

\noindent This example illustrates that ``most common'' translates to an aggregation with descending sort---a pattern that is obvious to a SQL expert but ambiguous to an LLM without guidance.

Together, these examples show how a diverse and representative set of question--data-model pairs teaches an LLM to map query patterns onto fragments of a data model, improving the accuracy and reliability of the SQL it generates. The challenge, as discussed in the introduction, is that creating such pairs manually is prohibitively expensive; the next section describes our approach for generating them automatically.

\begin{figure}[!t]
  \includegraphics[width=0.475\textwidth]{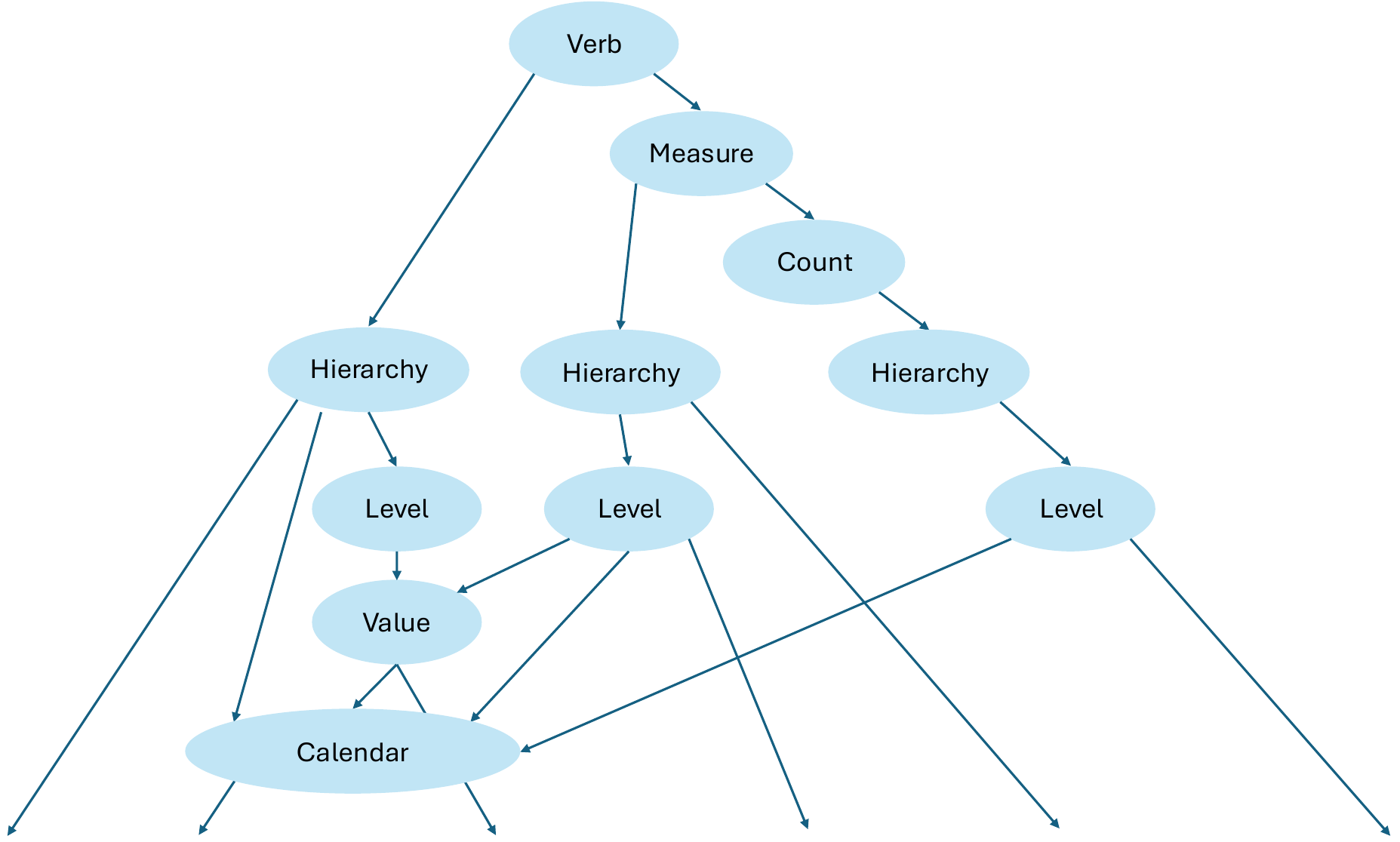}
  \caption{A seven-level query flow for constructing query context. Each path through the flow---from verb to calendar filter---defines a specific combination of operations and schema elements, producing a distinct data model and its corresponding natural language question.}
  \Description{A seven-level query flow showing verb, measures, counts, hierarchy, levels, value, and calendar/date filters}
  \label{fig:query-flows}
  \vspace{-0.2cm}
\end{figure}

\section{Automated Query Context}
\label{sec:approach}

We now describe our approach for automatically generating query context for a given database. The process involves three steps. First, we define a set of {\it query flows} that capture common patterns of data retrieval and analysis questions (\S\ref{sec:query-flows}). Second, we generate {\it data models} by systematically traversing these flows and instantiating them with appropriate measures, levels, filters, and post-operations (\S\ref{sec:generating-data-models}). Third, we {\it sample} specific schema elements and data values to populate these data models, using weighted strategies that ensure diversity and coverage (\S\ref{sec:sampling-elements}).

\begin{algorithm}[!t]
\caption{generate\_data\_models}
\label{label:QBGen.generate}
\SetAlgoLined
\KwIn{qb\_flows\_df: query flow definitions, config: hierarchy and measure configurations}
\KwOut{all\_models: list of generated data models}

Initialize empty list $all\_models$\;
\ForEach{flow row in qb\_flows\_df}{
  \ForEach{hierarchy in config}{
    Create base model with verb and hierarchy\;
    \If{flow is trend and no date columns exist}{
      Skip this hierarchy\;
    }
    \If{flow requires measures}{
      Sample measures using weighted selection\;
      \ForEach{sampled measure}{
        Attach aggregation functions (total, average, min, max)\;
        Apply 50\% probability weighting for ``total''\;
      }
    }
    \If{flow requires counts}{
      Add top/bottom N variants (top 1, 3, 10; bottom 3)\;
      Use different probability distributions for Compare vs Show\;
    }
    \If{flow requires levels}{
      Sample levels using inverse-frequency weighting\;
      \If{values needed}{
        Generate include/exclude filter combinations\;
      }
    }
    \If{flow requires calendar filters}{
      Generate date range combinations\;
      Generate date comparison combinations ($>$, $<$ a date)\;
      Attach calendar filters using round-robin distribution\;
    }
    Process post-operations: sort, project, having\;
    Generate natural language question via generate\_question()\;
  }
}
\Return{all\_models}\;
\end{algorithm}

\subsection{Query Flows}
\label{sec:query-flows}

The foundation of our approach is a set of {\it query flows} that represent common patterns of data retrieval and analysis questions. Each query flow maps a particular combination of relational operations (e.g., aggregation, filtering, sorting) to the types of schema elements (e.g., dimensions, measures) involved. By defining these flows upfront, we ensure that the generated query context covers a broad and systematic range of question types.

Figure~\ref{fig:query-flows} illustrates our query flows as a seven-level structure, where each level captures a different element of a question:

\begin{itemize}
    \item {\bf Verb}: The action word indicating the type of query (e.g., Show, List, Compare, Trend).
    \item {\bf Measures}: The quantitative metrics to be aggregated or analyzed (e.g., total sales, average revenue).
    \item {\bf Counts}: The number of top or bottom items to retrieve (e.g., top 10 products, bottom 3 regions).
    \item {\bf Hierarchy}: The primary table or entity being queried (e.g., sales, customers, orders).
    \item {\bf Levels}: The categorical dimensions for grouping or filtering (e.g., region, product category).
    \item {\bf Value}: The specific data values for filtering on levels (e.g., country=`Germany', product=`Laptop').
    \item {\bf Calendar/Date Filters}: Temporal constraints on the data (e.g., between Jan 2024 and June 2025, after March 2024).
\end{itemize}

A single path through the flow therefore yields one query context: one path might produce the question of Figure~\ref{fig:question-variations}, {\it ``Show total revenue for top-5 cities in Germany from Jan 2024 to June 2025''}, while another gives {\it ``Compare average order size by product category''}. Systematically navigating these paths constructs a wide variety of query contexts that map consistently onto the underlying data model structure.



\begin{algorithm}[!t]
\caption{sample\_measures}
\label{label:sample_measures} 
\SetAlgoLined
\KwIn{hierarchy: target hierarchy, postops: post-operations, measure\_limit: max measures to sample}
\KwOut{selected\_measures: list of sampled measures}

Get all measures for hierarchy from \_h\_measures\;
\If{no measures exist}{
  \Return{empty list}\;
}
\If{postops == ``having''}{
  \Return{level-1 measures only (aggregatable)}\;
}
Calculate sample limit: $limit \gets \min(total\_measures, 2 \text{ if postops else } measure\_limit)$\;
Fetch $measures\_level\_one$ (aggregatable) and $measures\_level\_two$ (non-aggregatable)\;
Calculate $level\_2\_quota \gets \min(available\_level\_2, limit - |measures\_level\_one|)$\;
Calculate $level\_1\_quota \gets limit - level\_2\_quota$\;
Random sample from each level according to quotas\;
\Return{level-1 samples + level-2 samples}\;
\end{algorithm}

\subsection{Generating Data Models}
\label{sec:generating-data-models}

Given the query flows defined above, the next step is to traverse them and produce concrete data models, instantiating each path with appropriate measures, levels, filters, and post-operations.

Algorithm~\ref{label:QBGen.generate} details this process. The outer loop iterates over each flow definition (e.g., Show with measures, Compare with counts) and each hierarchy (i.e., the primary table or entity). For each combination, it creates a base data model and progressively enriches it by: (i)~sampling measures and attaching aggregation functions such as total, average, min, and max (with a 50\% weighting toward ``total'' as the most common aggregation); (ii)~adding top/bottom-$N$ count variants where applicable; (iii)~sampling categorical levels for grouping or filtering, using inverse-frequency weighting to favor underrepresented columns; (iv)~generating date range and date comparison filters for temporal constraints; and (v)~applying post-operations such as sorting, projection, and HAVING-clause filters. Finally, the algorithm generates a natural language question for each completed data model using curated templates, and returns the full collection of data-model/question pairs.

\subsection{Sampling Elements}
\label{sec:sampling-elements}

Algorithm~\ref{label:QBGen.generate} relies on sampling specific measures and levels to populate each data model. A naive approach, such as uniform random sampling, would over-represent common columns and under-represent rare but important ones. We therefore use tailored sampling strategies for measures and levels, each designed to maximize diversity across the generated query contexts.

\smallskip\noindent{\bf Sampling measures.}
Algorithm~\ref{label:sample_measures} describes how we select measures for a given hierarchy. Measures fall into two categories: {\it level-1} (aggregatable) measures that support aggregation functions such as sum, average, min, and max (e.g., revenue, quantity), and {\it level-2} (non-aggregatable) measures that represent raw or pre-computed metrics (e.g., profit margin, satisfaction score). The algorithm prioritizes level-1 measures, since they produce the most common query patterns, while reserving a quota for level-2 measures to ensure broader coverage. For HAVING-clause post-operations, only level-1 measures are returned, since HAVING requires aggregate functions. When post-operations are present, the sampling limit drops to two measures to keep the resulting questions focused; otherwise it defaults to a configurable limit, typically four.

\begin{algorithm}[!t]
\caption{sample\_levels}
\label{label:sample_levels} 
\SetAlgoLined
\KwIn{hierarchy: target hierarchy, postops: post-operations, sample\_count: desired samples, needs\_value: whether values required}
\KwOut{sampled\_levels: list of selected hierarchy levels}

Get all levels for hierarchy from \_level\_counts\;
\ForEach{level in levels}{
  \If{level is date column OR all values are numeric}{
    Exclude from consideration\;
  }
  \If{needs\_value and level has no sample values}{
    Exclude from consideration\;
  }
}
\If{no valid levels remain}{
  \Return{empty list}\;
}
\If{postops == ``having''}{
  \Return{all valid levels}\;
}
Calculate inverse-frequency probabilities: $p_i \gets 1.0 / usage\_count_i$\;
Normalize probabilities to sum to 1.0\;
\If{sample\_count $<$ 0.25 $\times$ available levels}{
  $sample\_count \gets 2 \times sample\_count$\;
}
$sample\_count \gets \min(sample\_count, |valid\_levels|)$\;
Weighted random sample using probabilities\;
\ForEach{selected level}{
  \If{level ends in ``number'', ``count'', or ``id''}{
    $usage\_count[level] \gets usage\_count[level] + 10$\;
  }\Else{
    $usage\_count[level] \gets usage\_count[level] + 1$\;
  }
}
\Return{sampled\_levels}\;
\end{algorithm}

\smallskip\noindent{\bf Sampling levels.}
Algorithm~\ref{label:sample_levels} describes how we select hierarchy levels (i.e., categorical columns used for grouping or filtering). Not all columns in a hierarchy are suitable as levels: date columns are handled separately via calendar filters, purely numeric columns (e.g., IDs, counts) make poor grouping dimensions, and columns with no sample values cannot be used for filter examples. The algorithm first prunes these ineligible columns, then assigns each remaining level an {\it inverse-frequency} selection probability---levels that have been selected less frequently in prior iterations receive higher probability, creating a self-balancing mechanism that promotes diversity over time. To further discourage repetitive selections, columns ending in ``number'', ``count'', or ``id'' receive a 10$\times$ usage penalty when selected, since they typically produce less meaningful groupings. If the requested sample count would cover less than 25\% of the available levels, the algorithm doubles it to improve coverage.

\smallskip\noindent{\bf Adding metric filters.}
Algorithm~\ref{label:add_metric_filters} handles HAVING-clause style filters, which constrain aggregated values (e.g., ``total sales $>$ 5000''). For each aggregatable measure, the algorithm randomly selects an aggregate function, a comparator ($>$, $\geq$, $<$, $\leq$), and a threshold value. When the number of available measures is small ($\leq 5$), it generates two filters per measure (with different aggregates) to increase variety; otherwise, it generates one filter per measure. The resulting filters are attached to data model templates using a {\it greedy diversification} strategy that minimizes overlap on the measure and level dimensions, ensuring that the generated query contexts exhibit diverse combinations rather than repeating similar patterns.

\begin{algorithm}[!t]
\caption{add\_metric\_filters}
\label{label:add_metric_filters} 
\SetAlgoLined
\KwIn{flow: query flow definition, templates: current templates, hierarchy: target hierarchy}
\KwOut{(having\_template, updated\_templates)}

\If{flow has no level or templates is empty}{
  \Return{(None, templates)}\;
}
Get level-1 measures (aggregatable) for hierarchy\;
\If{no measures found}{
  \Return{(None, templates)}\;
}
\If{measure count $\leq$ 5}{
  $mode \gets$ ``double''\;
}\Else{
  $mode \gets$ ``single''\;
}
\ForEach{measure}{
  Get possible aggregates from config\;
  Randomly select aggregate\;
  Randomly select comparator from $\{>, \geq, <, \leq\}$\;
  Generate random threshold between 1 and 10,000\;
  Create filter: $\{aggregate, measure, comparator, threshold\}$\;
  \If{mode == ``double''}{
    Select different aggregate from remaining\;
    Generate new comparator and threshold\;
    Create second filter\;
  }
}
Apply filters to templates using greedy diversification on [measure, level]\;
\Return{(``With [metric]'', updated\_templates)}\;
\end{algorithm}

\section{Deployment Experience}
\label{sec:deployment}

Automated query context generation has been part of Tursio deployments for over a year, and included in its stable release since November 2025. Based on more than 50 such deployments and extensive internal testing, we find that a query context of at most 10,000 question--data-model pairs is sufficient to guide the LLM, even for large enterprise schemas; beyond this size, the additional pairs largely restate query patterns that are already covered. Generating a context of this size takes just 2--3 minutes end to end, which includes embedding the generated pairs into an internal pgvector store for semantic retrieval. Because generation is this inexpensive, the query context can be rebuilt as part of routine training rather than curated by hand, and it stays aligned with the database as schemas and data distributions evolve---in contrast to manually maintained sample repositories, which drift as soon as the underlying database changes.

Our deployment experience surfaced three further observations. First, and most importantly, the approach removes the cold start problem: unlike systems such as Genie~\cite{databricks-ai-bi-genie} and Cortex~\cite{snowflake-cortex-analyst}, which expect users to supply sample queries before the interface becomes useful, a new database can be onboarded with no human-authored examples at all. Second, the questions in the query context do not need to be polished English sentences. Template-generated phrasings such as {\it ``Show total revenue by city where country = Germany''} prove as effective for grounding the LLM as fluent, hand-written questions, since their role is to convey the mapping between query patterns and data-model fragments rather than to be read by a person. This lets us generate the natural language side of each pair from templates instead of paying for an additional LLM rewriting pass over the entire context. Third, beyond diversity, {\it composability} matters: because each generated pair exposes a small data-model fragment rather than a complete query, the LLM can combine fragments drawn from several retrieved pairs to answer questions that no single pair covers. This both reduces query generation errors and improves robustness on question phrasings that fall outside the generated flows.







\section{Related Work}
\label{sec:related-work}

\smallskip\noindent{\bf Natural language interfaces to databases.}
Beyond the query-by-example work discussed in \S\ref{sec:background}, a long line of systems has translated natural language directly into SQL. NaLIR~\cite{nalir} parses questions into a linguistic tree and interactively resolves ambiguities with the user, while ATHENA~\cite{athena} translates questions into an intermediate ontology query language before compiling to SQL, relying on a hand-built domain ontology. These systems established that a semantic description of the database---beyond the raw schema---is what makes accurate translation possible; our work can be seen as automatically generating such a description in the form of question--data-model pairs, rather than requiring it to be authored by a domain expert. The subsequent shift to learned and LLM-based translation is tracked by benchmarks such as Spider~\cite{spider_benchmark}, BIRD~\cite{bird_bench}, and BEAVER~\cite{beaver_benchmark}, the last of which shows that accuracy on academic benchmarks degrades sharply on real enterprise schemas---precisely the setting our deployments target.

\smallskip\noindent{\bf Context construction and example selection.}
A large body of recent work improves text-to-SQL accuracy by controlling what goes into the prompt. DIN-SQL~\cite{dinsql} decomposes generation into sub-tasks with schema-linking and self-correction stages; DAIL-SQL~\cite{dailsql} systematically studies question representation, example selection, and example organization, and shows that selecting demonstrations by combined question and query similarity is the dominant factor; and CHESS~\cite{chess} retrieves relevant values and prunes large schemas down to a context that fits the model. All of these methods {\it select} from an existing pool of examples or schema elements. Our work is complementary and addresses the prior question: where the pool comes from in the first place. Given a database with no query history and no curated examples, we generate the candidate pairs that such selection methods then rank and retrieve.

\smallskip\noindent{\bf Synthesizing text-to-SQL data.}
Synthetic generation has been used extensively to build {\it training} corpora. OmniSQL~\cite{omnisql} synthesizes SynSQL-2.5M, a million-scale cross-domain dataset spanning over 16,000 synthetic databases, and fine-tunes open models on it. Such approaches optimize a model's general text-to-SQL ability across many databases, and typically generate the databases along with the queries. Our objective is different: we generate context for one specific, existing production database, and consume it at inference time through retrieval rather than through fine-tuning. This distinction matters in practice: the generated pairs must reference real columns and real data values, must be regenerated as the schema evolves, and must remain small enough to retrieve from---which is what motivates the coverage-driven sampling strategies of \S\ref{sec:sampling-elements}.

\smallskip\noindent{\bf Mining query logs.}
An alternative source of examples is the workload itself. SnipSuggest~\cite{snipsuggest} recommends SQL snippets for each clause of a partially written query by mining a log of past queries, and SeeDB~\cite{seedb} searches the space of possible aggregate views to recommend interesting visualizations. LinkedIn's SQL Bot~\cite{linkedin-sql-bot} follows the same intuition at enterprise scale, building a knowledge graph from metadata, query logs, wikis, and code to retrieve context for its agent. These approaches are effective once a representative workload exists, but they cannot bootstrap a newly onboarded database---exactly the cold start problem we target. Our generation is driven by schema and data statistics alone, and can therefore run before the first user question arrives. The two combine naturally, with mined workload patterns reinforcing the generated context over time.

\smallskip\noindent{\bf Semantic layers and curated examples.}
The idea of a curated layer between raw tables and business questions predates LLMs: LookML~\cite{looker-lookml}, the dbt Semantic Layer~\cite{dbt-semantic-layer}, and Cube~\cite{cube-semantic-layer} let teams define measures, dimensions, and hierarchies once and reuse them across queries. Our notion of hierarchies, levels, and measures deliberately mirrors this vocabulary, but the artifacts we produce are examples for an LLM rather than definitions for a BI tool. Natural language interfaces layered on top of these models converge on the same mechanism---a repository of exemplar question--query pairs---and, almost universally, expect it to be filled in by hand: Databricks Genie through sample SQL queries and a knowledge store~\cite{databricks-ai-bi-genie, databricks-genie-knowledge-store}, Snowflake Cortex Analyst through semantic views and a Verified Query Repository~\cite{snowflake-cortex-analyst, snowflake-views-semantic-overview, snowflake-cortex-verified-query-repository}, Omni through {\it endorsed queries} that agents replicate~\cite{omni-context-modeling}, Timbr through a knowledge base of {\it approved examples}, i.e., vetted question--SQL pairs in which every approved query becomes future context~\cite{timbr-knowledge-base}, and ThoughtSpot~\cite{thoughtspot-spotter} and Vanna~\cite{vanna-ai-training} through analogous training mechanisms. Uber's QueryGPT~\cite{uber-querygpt} organizes curated tables and SQL samples into per-domain ``workspaces'' that an intent classifier routes questions to. This convergence is itself evidence for our formulation of query context; what differs is provenance. In each of these systems the pairs are human-authored assets whose curation cost scales with the schema and whose freshness decays as it evolves---the burden our generator removes.

\smallskip\noindent{\bf AI-native analytics platforms.}
A recent wave of systems attacks the same reliability problem from two other directions. One line constrains {\it how} the query is produced: TextQL compiles questions deterministically through an ontology and does not use the language model to write SQL directly~\cite{textql-ontology}, and Hasura's PromptQL has the model emit a multi-step plan in a DSL that is then executed deterministically~\cite{promptql}. This is complementary to our approach and consistent with our observation in \S\ref{sec:deployment} that composing answers from smaller, well-formed fragments reduces error. A second line, closest in spirit to our work, builds context automatically rather than asking users to write it: Myriade reverse-engineers the warehouse to map tables, validate metrics, and document relationships~\cite{myriade}; Genloop maintains a living context graph refined through a feedback loop over past interactions~\cite{genloop}; WisdomAI accumulates an adaptive context engine over the enterprise data stack~\cite{wisdomai}; and Google reports similar gains from deepening the model's understanding of the database~\cite{google-gemini-database-understanding}. These systems derive context from what already exists---schema profiling, query history, analyst feedback---which makes coverage a function of prior activity. Our generator is instead proactive: it enumerates the query space directly from the schema and data distribution, so coverage is systematic by construction and available before the first user question.

\begin{acks}
We thank both our current and former colleagues at Tursio for their contributions to building and refining the query context generator, and our on-premises customers for their valuable feedback throughout the deployment process.
\end{acks}

\balance
\bibliographystyle{ACM-Reference-Format}
\bibliography{references}

@String{Chelsea = "Chelsea" }

@String{Springer = "Springer-Verlag" }

@inproceedings{bird_bench,
 author = {Li, Jinyang and Hui, Binyuan and Qu, Ge and Yang, Jiaxi and Li, Binhua and Li, Bowen and Wang, Bailin and Qin, Bowen and Geng, Ruiying and Huo, Nan and Zhou, Xuanhe and Chenhao, Ma and Li, Guoliang and Chang, Kevin and Huang, Fei and Cheng, Reynold and Li, Yongbin},
 booktitle = {Advances in Neural Information Processing Systems},
 title = {Can LLM Already Serve as A Database Interface? A BIg Bench for Large-Scale Database Grounded Text-to-SQLs},
 year = {2023}
}

@misc{beaver_benchmark,
      title={BEAVER: An Enterprise Benchmark for Text-to-SQL}, 
      author={Peter Baile Chen and Fabian Wenz and Yi Zhang and Devin Yang and Justin Choi and Nesime Tatbul and Michael Cafarella and Çağatay Demiralp and Michael Stonebraker},
      year={2025},
      archivePrefix={arXiv},
      url={https://arxiv.org/abs/2409.02038}, 
}

@online{databricks-genie-knowledge-store,
  author = {Databricks},
  title = {Databricks Genie Knowledge Store},
  year = {2025},
  url = {https://docs.databricks.com/aws/en/genie/knowledge-store},
  lastaccessed = {October 7, 2025},
}

@online{snowflake-views-semantic-overview,
  author = {Snowflake},
  title = {Overview of Semantic Views},
  year = {2025},
  url = {https://docs.snowflake.com/en/user-guide/views-semantic/overview},
  lastaccessed = {November 15, 2025},
}

@online{snowflake-cortex-verified-query-repository,
  author = {Snowflake},
  title = {Cortex Analyst Verified Query Repository},
  year = {2025},
  url = {https://docs.snowflake.com/en/user-guide/snowflake-cortex/cortex-analyst/verified-query-repository},
  lastaccessed = {November 20, 2025},
}

@online{google-gemini-database-understanding,
  author = {Google Cloud},
  title = {A new top score: Advancing Text-to-SQL on the BIRD benchmark},
  year = {2025},
  lastaccessed = {November 14, 2025},
  url = {https://cloud.google.com/blog/products/databases/how-to-get-gemini-to-deeply-understand-your-database}
}

@online{databricks-ai-bi-genie,
  author = {Databricks},
  title = {Databricks AI/BI Genie},
  year = {2026},
  url = {https://docs.databricks.com/aws/en/ai-bi/},
  lastaccessed = {January 6, 2026},
}

@online{snowflake-cortex-analyst,
  author = {Snowflake},
  title = {Snowflake Cortex Analyst},
  year = {2026},
  url = {https://docs.snowflake.com/en/user-guide/snowflakecortex/cortex-analyst},
  lastaccessed = {January 6, 2026},
}

@online{microsoft-fabric-copilot,
  author = {Microsoft},
  title = { Microsoft Fabric Copilot},
  year = {2026},
  lastaccessed = {January 6, 2026},
  url = {https://learn.microsoft.com/enus/fabric/fundamentals/copilot-fabric-overview}
}

@online{thoughtspot-spotter,
  author = {ThoughtSpot},
  title = {ThoughtSpot Spotter},
  year = {2026},
  url = {https://docs.thoughtspot.com/cloud/10.15.0.cl/spotter},
  lastaccessed = {January 6, 2026},
}

@online{vanna-ai-training,
  author = {Vanna AI},
  title = {Vanna AI Training Documentation},
  year = {2026},
  url = {https://vanna.ai/docs/placeholder/training},
  lastaccessed = {January 6, 2026},
}

@inproceedings{DBLP:conf/vldb/Zloof75,
  author       = {Mosh{\'{e}} M. Zloof},
  title        = {Query-by-Example: the Invocation and Definition of Tables and Forms},
  booktitle    = {VLDB},
  pages        = {1--24},
  publisher    = {{ACM}},
  address      = {}, 
  year         = {1975},
}

@ARTICLE{5388055,
  author={Mosh{\'{e}} M. Zloof},
  journal={IBM Systems Journal}, 
  title={Query-by-Example: A data base language}, 
  year={1977},
  volume={16},
  number={4},
  pages={324-343},
}

@book{exampleBasedMethods,
  title     = "Data Exploration Using Example-Based Methods",
  author    = "Matteo Lissandrini and Davide Mottin and Themis Palpanas and Yannis Velegrakis",
  year      = 2019,
  edition   = 1,
  publisher = "Springer Cham",
  address   = "Switzerland"
}

@article{10.14778/2831360.2831369,
  author = {Li, Hao and Chan, Chee-Yong and Maier, David},
  title = {Query from examples: an iterative, data-driven approach to query construction},
  year = {2015},
  volume = {8},
  number = {13},
  journal = {PVLDB},
  pages = {2158--2169},
}

@inproceedings{10.1145/2588555.2593664,
  author = {Shen, Yanyan and Chakrabarti, Kaushik and Chaudhuri, Surajit and Ding, Bolin and Novik, Lev},
  title = {Discovering queries based on example tuples},
  year = {2014},
  booktitle = {SIGMOD},
  pages = {493--504},
  publisher = {ACM},
  address = {New York, NY, USA},
}

@inproceedings{ai-machine,
  author = {Alekh Jindal and Shi Qiao and Sathwik Reddy Madhula and Kanupriya Raheja and Sandhya Jain},
  title = {Turning Databases Into Generative AI Machines},
  year = {2024},
  booktitle = {CIDR},
  publisher = {CIDR},
  address = {Chaminade, USA},
}

@online{cursor-ai,
  author = {Anysphere},
  title = {Cursor -- The AI Code Editor},
  year = {2026},
  url = {https://www.cursor.com/},
  lastaccessed = {February 19, 2026},
}

@online{lovable-ai,
  author = {Lovable},
  title = {Lovable -- Your superhuman full stack engineer},
  year = {2026},
  url = {https://lovable.dev/},
  lastaccessed = {February 19, 2026},
}

@online{bolt-ai,
  author = {StackBlitz},
  title = {Bolt.new -- Prompt, run, edit, and deploy full-stack web apps},
  year = {2026},
  url = {https://bolt.new/},
  lastaccessed = {February 19, 2026},
}

@online{replit-ai,
  author = {Replit},
  title = {Replit Agent -- Build apps with AI},
  year = {2026},
  url = {https://replit.com/},
  lastaccessed = {February 19, 2026},
}

@inproceedings{spider_benchmark,
  author    = {Yu, Tao and Zhang, Rui and Yang, Kai and Yasunaga, Michihiro and Wang, Dongxu and Li, Zifan and Ma, James and Li, Irene and Yao, Qingning and Roman, Shanelle and Zhang, Zilin and Radev, Dragomir},
  title     = {Spider: A Large-Scale Human-Labeled Dataset for Complex and Cross-Domain Semantic Parsing and Text-to-SQL Task},
  booktitle = {Proceedings of the 2018 Conference on Empirical Methods in Natural Language Processing (EMNLP)},
  pages     = {3911--3921},
  year      = {2018},
  address   = {Brussels, Belgium},
}

@article{dailsql,
  author  = {Gao, Dawei and Wang, Haibin and Li, Yaliang and Sun, Xiuyu and Qian, Yichen and Ding, Bolin and Zhou, Jingren},
  title   = {Text-to-{SQL} Empowered by Large Language Models: A Benchmark Evaluation},
  journal = {PVLDB},
  volume  = {17},
  number  = {5},
  pages   = {1132--1145},
  year    = {2024},
  doi     = {10.14778/3641204.3641221},
}

@inproceedings{dinsql,
  author    = {Pourreza, Mohammadreza and Rafiei, Davood},
  title     = {{DIN-SQL}: Decomposed In-Context Learning of Text-to-{SQL} with Self-Correction},
  booktitle = {Advances in Neural Information Processing Systems (NeurIPS)},
  year      = {2023},
}

@inproceedings{chess,
  author    = {Talaei, Shayan and Pourreza, Mohammadreza and Chang, Yu-Chen and Mirhoseini, Azalia and Saberi, Amin},
  title     = {{CHESS}: Contextual Harnessing for Efficient {SQL} Synthesis},
  booktitle = {Proceedings of the 42nd International Conference on Machine Learning (ICML)},
  year      = {2025},
  url       = {https://arxiv.org/abs/2405.16755},
}

@article{omnisql,
  author  = {Li, Haoyang and Wu, Shang and Zhang, Xiaokang and Huang, Xinmei and Zhang, Jing and Jiang, Fuxin and Wang, Shuai and Zhang, Tieying and Chen, Jianjun and Shi, Rui and Chen, Hong and Li, Cuiping},
  title   = {{OmniSQL}: Synthesizing High-Quality Text-to-{SQL} Data at Scale},
  journal = {PVLDB},
  volume  = {18},
  number  = {11},
  pages   = {4695--4709},
  year    = {2025},
  doi     = {10.14778/3749646.3749723},
}

@misc{linkedin-sql-bot,
  author        = {Chen, Albert and Bundele, Manas and Ahlawat, Gaurav and Stetz, Patrick and Wang, Zhitao and Fei, Qiang and Jung, Donghoon and Chu, Audrey and Jayaraman, Bharadwaj and Panth, Ayushi and Arora, Yatin and Jain, Sourav and Varma, Renjith and Ilin, Alexey and Melnychuk, Iuliia and Chueh, Chelsea and Sil, Joyan and Wang, Xiaofeng},
  title         = {Text-to-{SQL} for Enterprise Data Analytics},
  year          = {2025},
  archivePrefix = {arXiv},
  url           = {https://arxiv.org/abs/2507.14372},
}

@article{snipsuggest,
  author  = {Khoussainova, Nodira and Kwon, YongChul and Balazinska, Magdalena and Suciu, Dan},
  title   = {{SnipSuggest}: Context-Aware Autocompletion for {SQL}},
  journal = {PVLDB},
  volume  = {4},
  number  = {1},
  pages   = {22--33},
  year    = {2010},
  doi     = {10.14778/1880172.1880175},
}

@article{nalir,
  author  = {Li, Fei and Jagadish, H. V.},
  title   = {Constructing an Interactive Natural Language Interface for Relational Databases},
  journal = {PVLDB},
  volume  = {8},
  number  = {1},
  pages   = {73--84},
  year    = {2014},
  doi     = {10.14778/2735461.2735468},
}

@article{athena,
  author  = {Saha, Diptikalyan and Floratou, Avrilia and Sankaranarayanan, Karthik and Minhas, Umar Farooq and Mittal, Ashish R. and {\"O}zcan, Fatma},
  title   = {{ATHENA}: An Ontology-Driven System for Natural Language Querying over Relational Data Stores},
  journal = {PVLDB},
  volume  = {9},
  number  = {12},
  pages   = {1209--1220},
  year    = {2016},
  doi     = {10.14778/2994509.2994536},
}

@article{seedb,
  author  = {Vartak, Manasi and Rahman, Sajjadur and Madden, Samuel and Parameswaran, Aditya and Polyzotis, Neoklis},
  title   = {{SeeDB}: Efficient Data-Driven Visualization Recommendations to Support Visual Analytics},
  journal = {PVLDB},
  volume  = {8},
  number  = {13},
  pages   = {2182--2193},
  year    = {2015},
  doi     = {10.14778/2831360.2831371},
}

@online{uber-querygpt,
  author       = {Uber},
  title        = {{QueryGPT} -- Natural Language to {SQL} Using Generative {AI}},
  year         = {2024},
  url          = {https://www.uber.com/blog/query-gpt/},
  lastaccessed = {August 6, 2026},
}

@online{dbt-semantic-layer,
  author       = {dbt Labs},
  title        = {{dbt} Semantic Layer},
  year         = {2025},
  url          = {https://docs.getdbt.com/docs/use-dbt-semantic-layer/dbt-sl},
  lastaccessed = {August 6, 2026},
}

@online{looker-lookml,
  author       = {Google Cloud},
  title        = {What is {LookML}?},
  year         = {2025},
  url          = {https://cloud.google.com/looker/docs/what-is-lookml},
  lastaccessed = {August 6, 2026},
}

@online{cube-semantic-layer,
  author       = {Cube Dev},
  title        = {Cube: The Universal Semantic Layer},
  year         = {2025},
  url          = {https://cube.dev/},
  lastaccessed = {August 6, 2026},
}

@online{omni-context-modeling,
  author       = {Omni Analytics},
  title        = {Context Modeling},
  year         = {2026},
  url          = {https://omni.co/context-modeling},
  lastaccessed = {August 6, 2026},
}

@online{timbr-knowledge-base,
  author       = {Timbr},
  title        = {Knowledge Base for Data Agents},
  year         = {2026},
  url          = {https://timbr.ai/timbr-core/knowledge-base-for-data-agents/},
  lastaccessed = {August 6, 2026},
}

@online{textql-ontology,
  author       = {TextQL},
  title        = {Ontology},
  year         = {2026},
  url          = {https://textql.com/products/ontology},
  lastaccessed = {August 6, 2026},
}

@online{promptql,
  author       = {Hasura},
  title        = {{PromptQL}: Reliable {AI} on Any Data},
  year         = {2026},
  url          = {https://hasura.io/promptql},
  lastaccessed = {August 6, 2026},
}

@online{myriade,
  author       = {Myriade},
  title        = {Reliable {AI} Analytics, Even on Messy Data},
  year         = {2026},
  url          = {https://www.myriade.ai/},
  lastaccessed = {August 6, 2026},
}

@online{genloop,
  author       = {Genloop},
  title        = {Genloop: Context-Aware Data Intelligence},
  year         = {2026},
  url          = {https://genloop.ai/},
  lastaccessed = {August 6, 2026},
}

@online{wisdomai,
  author       = {WisdomAI},
  title        = {Agentic Analytics for Enterprises},
  year         = {2026},
  url          = {https://www.wisdom.ai/},
  lastaccessed = {August 6, 2026},
}

\end{document}